\documentclass[12pt,a4paper,english]{article}
\usepackage[ansinew]{inputenc}
\usepackage{babel}
\usepackage{latexsym}
\usepackage{amsmath}
\usepackage{amsfonts}
\usepackage{amssymb}
\usepackage{textcomp}
\usepackage{graphicx}
\usepackage{enumerate}

\begin{document}

\begin{titlepage}
\begin{center}
{\large A perturbation approach to Translational Gravity}

\vglue 1cm

J. JULVE* and A. TIEMBLO$^+$

{\it *$^+$Instituto de Física Fundamental, CSIC, \break C/Serrano
113 bis, Madrid 28006, Spain

*julve@iff.csic.es

 $^+$tiemblo@iff.csic.es}

\end{center}

\vglue 1cm

\noindent {Within} a gauge formulation of 3+1 gravity relying on
a nonlinear realization of the group of isometries of space-time,
a natural expansion of the metric tensor arises and a simple choice
of the gravity dynamical variables is possible. We show that the
expansion parameter can be identified with the gravitational constant
and that the first order depends only on a diagonal matrix
in the ensuing perturbation approach. The explicit first order 
solution is calculated in the static isotropic case, and its general 
structure is worked out in the harmonic gauge.

\vglue 0.5cm

\noindent{\it Keywords} : Nonlinear realizations; gauge theory of
gravitation; gauge translations; minimal
tetrads.

\vglue 1cm

\begin{center}
{\it PACS} : 04.20.Cv , 04.20.Fy , 02.20.Sv
\end{center}

\vglue 3cm

\noindent{*Corresponding author.}

\end{titlepage}

\section{Introduction}

A basic issue of the theories of gravity is the adoption of suitable
variables. The advent of Relativity replaced the historic Newtonian
gravitational field and potential by a geometrical description in
which the metric tensor has the dominant role. Later on, other
variables related to it have been considered as, for instance, the
Vierbein, Lapse-Shift-3Metric \cite{Arnowitt}\cite{Misner},
Ashtekar's \cite{Ashtekar}, loop variables, etc.
They emphasize different physical aspects, respectively the relationship
of general coordinates to local Lorentzian coordinates, the canonical
Hamiltonian formulation and the more recent self-dual action, etc.
Lastly, we quote the gauge point of view
\cite{Utiyama}\cite{Sciama}\cite{Kibble}\cite{Eguchi}\cite{Choquet},
which embodies some of the features of the above and adds interesting
new possibilities.

The gauge field formulation of gravity, see for instance
\cite{Hehl1}\cite{Hehl2}\cite{Hehl3},
has deserved attention since a long time for several reasons.
First, it makes gravity more amenable to a unified field description
with all the other fundamental interactions, and, no less important,
it allows the coupling to fermion fields.
In all of these properties the vierbein has a key role.

In fact, a unifying gauge scheme requires finding a suitable gauge
connection mediating gravity. To this purpose, the tetrads, featuring
holonomic and nonholonomic indices, prove to be essential. On the
other hand, the evidence that elementary matter is fermionic strongly
supports the hypothesis that gravity couples to it through the vierbein,
which entails enlarging the geometrical framework of General Relativity
with the introduction of a suitable internal group
\cite{Hayashi}\cite{Ivanenko}\cite{Lord1}\cite{Lord2}\cite{Sardanashvily}.
The assumption that the vierbein depends on the connection of the local
translations makes it to transform as a tensor under diffeomorphisms
and under the (even local) Lorentz group. These properties can be
implemented by defining the vierbein by means of a Non Linear
Realization (NLR)
\cite{Coleman}\cite{Callan}\cite{Salam}\cite{Isham}\cite{Borisov}
of the Poincar\'e group (cosets with respect to Lorentz).

The establishment of a gauge symmetry relies on the empirical
evidence of a locally invariant property related to a group of
transformations. The existence of a continuous ten-parameter group
(Poincar\'e) behind the conservation of the fundamental dynamical
variables, strongly suggests the existence of a relevant link
between dynamics and the basic properties of the space-time i.e
geometry. From the geometrical point of view Poincar\'e's may be
defined as the group of isometries of Minkowsky space, so that it
seems most natural to consider the group of isometries of that space
as the gauge group of a dynamical theory of the space-time.

Minkowsky's is the simplest case (zero curvature) of a maximally
symmetric 3+1 space and thus it excludes the presence of a
cosmological constant. On the other hand, our knowledge of the
geometrical properties of the space-time is only phenomenological
and therefore it is approximate. Strictly speaking what we locally
observe is that the space is approximately homogeneous and
isotropic, and that it is endowed with the kinematical Lorentz group
of Relativity so that we can assume that the global symmetry group
of space-time seems to be locally very close to Poincar\'e's.
Consequently we assume that the general candidate for a gauge theory
of gravity is the group of isometries of a maximally symmetric space
(the limit of zero curvature being the Poincar\'e group).

A close study of this gauge theory shows the possibility of
adopting highly convenient dynamical variables that let splitting
the ten traditional gravitational variables into the four components
of a diagonal matrix and six antisymmetric tensor components.
This is the main contents of this paper. As a consequence, on very
general grounds, these variables let to introduce a characteristic
length $\lambda$ on which a useful perturbation expansion can be
worked out \cite{Julve3}. We remark that the corresponding splitting
of the vierbein entails a truncated expansion of the metric tensor
which is of a different nature from the traditional decomposition
into a flat plus a perturbation metric, usually adopted in the weak
field approximation.

To make the paper most self contained we outline in Section 2 the
main features of the non linear local realizations of the space-time
groups introduced in earlier works
\cite{Julve1}\cite{Julve2}\cite{Lopezpinto}\cite{Lopezpinto2}\cite{Tresguerres}\cite{Martin1}\cite{Martin2}.

In Section 3 we compute the first terms in $\lambda$  of the solutions
to the field equations for a static and isotropic metric, the first order yielding
directly the post-Newtonian potential. This result reveals a relationship
between $\lambda$ and the gravitational constant.

Section 4 addresses the expansion of the theory to first order in $\lambda$,
the basic building block for the computation of the higher order terms of
the general solution.

Finally the the conclusions are drawn in Section 5 together with an outline of some
open problems.

\section{Non linear gauge realizations of space-time symmetry}

We briefly review here some fundamental tools and results from
previous works \cite{Julve3}.

Let $G$ be a Lie group having a subgroup $H$, we assume that the
elements $C(\varphi)$ (cosets) of the quotient space $G/H$ can be
characterized by a set of parameters say $\varphi$. Let us denote by
$\psi$ an arbitrary linear representation of the subgroup $H$.

The non linear realization can be derived from the action of a
general element  $"g"$ of the whole group on the coset
representatives defined in the form:

\begin{equation}\label{1}
g\,C(\varphi)=C(\varphi')h(\varphi,g)
\end{equation}
where
$h(\varphi,g)\in H$. It acts linearly on the representation space
$\psi$ according to:

\begin{equation}\label{2}
\Psi'=\varrho[h(\varphi,g)]\,\Psi,
\end{equation}
being $\varrho[h]$ a representation of the subgroup $H$.

The next step to construct a theory with non linear local symmetry
is to define suitable gauge connections. They can be obtained by
substituting the ordinary Cartan 1-form $\omega=C^{-1}dC$ by a
generalized expression of the form:

\begin{equation}\label{3}
\Gamma=C^{-1}{\bf D}C
\end{equation}
where $D=d+\Omega$ is the covariant differential built with the
1-form connection $\Omega$ defined on the algebra of the whole group
and having the canonical transformation law:

\begin{equation}
\Omega'=g\Omega g^{-1}+  gdg^{-1}
\end{equation}

\noindent{Then} our generalized local Cartan 1-form is:
\begin{equation}
\Gamma=e^{-{\rm i}\,\varphi^iP_i}(d+{\rm i}\,T^iP_i+\frac{\rm
i}{2}A^{ij}L_{ij})e^{{\rm i}\,\varphi^iP_i}
\end{equation}
where $T^i$ is the linear translational connection, $A^{ij}$ is the
corresponding one for the Lorentz group, $P_i$ and $L_{ij}$ are
the generators of translations and Lorentz transformations respectively,
and we adopt the metric $\eta_{ij}=(\,1\,,1\,,1\,,-1)$.

In (5), the translational connection 1-form $T^i$ has dimensions of
length. In order to have a dimensionless connection $\gamma^i$
homogeneous with the ordinary Lorentz connection $A^{ij}$, we
introduce a constant characteristic length, say $\lambda$\,, and
define $T^i=\lambda\gamma^i$. Using Hausdorff-Campbell formulas
we obtain:

\begin{equation}
e^i=D\varphi^i+\lambda\gamma^i\;\equiv e(0)^i+\lambda\gamma^i\;,
\end{equation}
where
\begin{equation}
D\varphi^i=d\varphi^i+A^i_{\,j}\varphi^j\;
\end{equation}
is the Lorentz covariant derivative of the coset field $\varphi^i$ .
We observe here that the covariant derivative of a Lorentz vector
like $e(0)^i$ in (6) is the minimal structure suited to define
a tetrad, so we shall call it "minimal tetrad".
We stress that the difference between $e(0)^i$
and $e^i$ regards the behavior under local translations,
due to the presence of the connection $\gamma^i\,$.

The tetrad (6) gives rise to a metric tensor of
the form:
\begin{equation}
g_{\mu\nu}=e_{i\mu}e^i_\nu\,
 =g(0)_{\mu \nu}+\lambda\gamma_{(\mu
 \nu)}+\lambda^2\gamma_{\mu \rho}\gamma_{\nu\sigma}\,g(0)^{\rho
 \sigma}\;,
\end{equation}
where $g(0)_{\mu \nu}=e(0)_{i\mu}e(0)_\nu^i$ is the
corresponding "minimal metric tensor", and we have used $
e(0)_{i\mu}$ and its formal inverse $e(0)^\nu_j$ to
transform indices.

Two comments are now in order. The first one concerns equation(8)
that mimics a weak field expansion over a background metric
$g(0)_{\mu \nu}$. It must be emphasized however that it is not a
perturbation approach but an exact result derived from the
underlying gauge structure, which is apparent only at the vierbein
level. Secondly, the decomposition (8) implies a non-trivial
structure for the formal inverse $e^\mu_i$ present in the
definition of the contravariant metric tensor. We explicitly assume
that the theory is analytical in the characteristic length
$\lambda$, so that the formal inverses are given by an expansion in
powers of $\lambda$.

Now we are to show that the structure and properties of this minimal
metric tensor can be derived from general integrability conditions.
To this end we first redefine the Lorentz connection
$A_\mu^{ij}$ as follows:
\begin{equation}
A_\mu^{ij}=e^{\alpha i}{\cal D}_\mu
e_\alpha^j+B_\mu^{ij},
\end{equation}
where ${\cal D}_\mu$ is the ordinary Christoffel covariant
derivative acting on the coordinate index $\alpha$ of the tetrad
$e_\alpha^j$ . The first term of this redefinition, usual in
gauge theories of gravity, describes the value of the Lorentz
connection in the absence of matter. For $\lambda=0$, contracting
(9) with $\varphi^j$ we get
\begin{equation}
A_{\mu j}^i\varphi^j=e(0)^{\alpha i}{\cal
D}(0)_\mu[e(0)_{\alpha j}\varphi^j]-\partial_\mu\varphi^i+B_\mu^i\,,
\end{equation}
where ${\cal D}(0)_\mu$ is the Christoffel covariant derivative
constructed with the metric tensor $g(0)_{\mu \nu}$ , and
$B_\mu^i=B_{\mu j}^i\varphi^j$.

Taking into account (6) we get
\begin{equation}
e(0)_{\mu i}= e(0)^\alpha_i D_\mu D_\alpha\,\sigma+B_{\mu i}\,,
\end{equation}
where $\sigma\equiv\frac{1}{2}\eta_{ij}\,\varphi^i\varphi^j$.
Contracting (11) with $e(0)_\nu^i$ we finally have
\begin{equation}
g(0)_{\mu \nu}=D(0)_\mu D(0)_\nu\,\sigma+B_{\mu \nu}\,,
\end{equation}
where $B_{\mu\nu}=B_{\mu i}e_\nu^i$ . As a consequence of (12), the antisymmetric
part $B_{[\mu\nu]}$ vanishes. The symmetry properties of $B_{ijk}$ imply
\cite{Julve3}\cite{Martin1} that the symmetric part $B_{[\mu\nu]}$ vanishes as well,
and therefore
\begin{equation}
g(0)_{\mu \nu}=D(0)_\mu D(0)_\nu\,\sigma\,.
\end{equation}
Using the trace of (13), namely $4=\Box\sigma$, this equation becomes
\begin{equation}
D(0)_\mu D(0)_\nu\,\sigma=\frac{1}{4}g(0)_{\mu \nu}\Box\sigma\,.
\end{equation}
It is well known that the integrability condition for this equation has the
maximally symmetric spaces as solution, as expected.

We recall that (1) and (2) show that the fields $\varphi^i$, being the Goldstone
bosons of the gauged translations and thus isomorphic to the Cartesian coordinates,
are by definition independent functions. Taking for them these coordinates, (13)
yields
\begin{equation}
g(0)_{ij}=\eta_{\,ij}\,,
\end{equation}
with vanishing cosmological constant. In a maximally symmetric space, the Christoffel
symbols in the covariant derivatives in (13) are proportional to the sectional curvature
and therefore they vanish in the absence of a cosmological constant.

The passage from the gauge description to the geometrical one is
canonically accomplished by using (9) for $B_\mu^{ij}=0$ in the
Field Strength Tensor, which becomes:
\begin{equation}
F_{\mu\nu}^{ij}=e_\alpha^i e_\beta^j R^{\alpha\beta}_{\mu\nu}
\end{equation}

Starting from this relationship we first consider the Poincar\'e
case where the cancelation of the Riemann tensor at zero order in
$\lambda$ stems from the integrability conditions, so that, being
$R(0)^{\alpha\beta}_{\mu\nu}=0$, we conclude that also $F(0)_{\mu
\nu}^{ij}=0$ and then $A_\mu^{ij}$ must either vanish or, more generally,
be a pure gauge. The structure of such a connection is given by the
inhomogeneous part of the formal variation of a gauge connection, thus we write:
\begin{equation}
A(0)_\mu^{ij}=U^{ik}\partial_\mu U_k^j,
\end{equation}
where $U^{ik}$ is an arbitrary pseudo-orthogonal matrix describing a
general Lorentz transformation. Putting this in the zeroth order of
(18) we obtain:
\begin{equation}
e(0)_\mu^i=\partial_\mu\varphi^i+U^{ik}\partial_\mu
U_{kj}\varphi^j\;,
\end{equation}
so we can write:
\begin{equation}
e(0)_\mu^i=\partial_\mu \varphi^i+
U^{ik}\partial_\mu[U_{kj}\varphi^j] -\partial_\mu
\varphi^i=U^{ik}\partial_\mu [U_{kj}\varphi^j] =
U^{ik}\partial_\mu{\hat\varphi}_k
\end{equation}
where ${\hat\varphi}_k=U_{kj}\varphi^j$. Then the background metric
may be written as follows:
\begin{equation}
g(0)_{\mu\nu}=U^{ik}\partial_\mu{\hat\varphi}_k U^{il}\partial_\nu
{\hat\varphi}_l=\partial_\mu{\hat\varphi}_k
\partial_\nu{\hat\varphi}^k
\end{equation}

Now we recover the expression (8) of the general metric tensor,
taking the symmetric and antisymmetric parts of $\gamma_{\mu \nu}$:
\begin{equation}
\gamma_{\mu \nu}=\frac{1}{2}\,s_{\mu \nu}+\frac{1}{2}\,a_{\mu \nu },
\end{equation}
being $s_{\mu  \nu}=\gamma_{(\mu \nu)}$ and $a_{\mu
\nu}=\gamma_{[\mu \nu]}$ , so (19) becomes:
\begin{equation}
g_{\mu \nu}=g(0)_{\mu \nu}+\lambda\,s_{\mu
\nu}+\frac{\lambda^2}{4}[s_{\mu\rho}s_{\nu \rho}+s_{(\mu
\rho}a_{\nu) \sigma}+a_{\mu \rho}a_{\nu \sigma}]g(0)^{\rho \sigma}.
\end{equation}
Steering to Lorentz indices, namely $s_{\mu
\nu}=e(0)_\mu^i\,s_{ij}\,e(0)_\nu^j$ and
$a_{\mu\nu}=e(0)_\mu^i\,a_{ij}\,e(0)_\nu^j$ , and adopting the
coordinates $x^\mu$ for the cartesian Goldstone ones $\varphi^i$,
the metric tensor $g(0)_{\mu\nu}$ reduces to $\eta_{i j}$ . Now we
can choose $U$ such that $s_{ij}$ becomes $U^k_is_{kl}U^l_j=d_{ij}$
diagonal, and $a_{ij}\rightarrow U^k_ia_{kl}U^l_j={\hat a}_{ij}$ ,
obtaining:
\begin{equation}
g_{i j}=\eta_{i j}+\lambda\,d_{i j}+\frac{\lambda^2}{4}[d_{i k}d_{j
l}+d_{(i k}\hat{a}_{j)l}+\hat{a}_{i k}\hat{a}_{j l}]\eta^{k l}.
\end{equation}

We then have the usual ten degrees of freedom of canonical gravity,
albeit in quite a different arrangement: the four eigenvalues of the
symmetric part of $\gamma_{\mu\nu}$ and the six elements of an
antisymmetric matrix. We stress that these d.o.f. appear in (23) at
different orders in $\lambda$ so that, being $d_{ij}$ diagonal, the
calculations at first order get highly simplified.

The case of a maximally symmetric space is slightly more complicated
because the generators of the translations do not commute, namely
\begin{equation}
[P_i\,,P_j]={\rm i}\,k L_{i j}\;.
\end{equation}
The result has the same form (23) albeit for the substitution of the
flat metric $\eta_{ij}$ by a maximally symmetric metric, as for instance
by its Riemannian form
\begin{equation}
g(0)_{ij}= (1+\frac{1}{4}\,{\chi})^{-2}\eta_{ij}\;,
\end{equation}
where $\chi\equiv k \,\eta_{ij}\,\varphi^i\varphi^j$ .
The details can be found in \cite{Julve3}\cite{Martin1}\cite{Martin2}.

\section{Static isotropic solution}

The case of the static and 3-space maximally symmetric solution
has the twofold interest of illustrating the workings of our approach
and also of providing a direct physical interpretation of the expansion
parameter $\lambda$ . The choice of the dynamical variables given in
equation (23) drastically simplifies the structure of the theory and it
lets us to work out the general form and properties of the vacuum equations
of gravity at lower orders in $\lambda$ . To this end we start from the
Einstein's equations in the absence of a cosmological constant,
namely $R_{ij}=0$ .

A useful formula relating the Christophel's connections $\Gamma^\varrho_{\mu\alpha}$
and $\Gamma(0)^\varrho_{\mu\alpha}$ , corresponding respectively to the two different
metric tensors $g_{\mu\nu}$ and $g(0)_{\mu\nu}$, is

\begin{equation}
\Gamma^\rho_{\mu\alpha}=\Gamma(0)^\rho_{\mu\alpha}+\Delta^\rho_{\mu\alpha}\;,
\end{equation}
where
\begin{equation}
\Delta^\rho_{\mu\alpha}=\frac{1}{2}g^{\lambda \rho}[D(0)_\mu
g_{\lambda\alpha}+D(0)_\alpha g_{\lambda \mu}-D(0)_\lambda g_{\mu
\alpha}]\;,
\end{equation}
being $D(0)_\mu$ the covariant derivative in terms of
$\Gamma(0)^\rho_{\mu\alpha}$.

Accordingly with our choice of coordinates we will use Latin
indexes in the following , being $g(0)_{ij}$ the background metric and
$g_{ij}$ given by (23).
The relation between the respective Ricci tensors is then
\begin{equation}
R_{ij}=R(0)_{ij}+D(0)_j\Delta^k_{ki}-D(0)_k\Delta^k_{ij}+\Delta^l_{ik}\Delta^k_{j\,l}-\Delta^l_{ij}\Delta^k_{kl},
\end{equation}
where $D(0)_i$ and $R(0)_{ij}$ are the covariant derivative and the Ricci
tensor stemming from the background metric.

For the Poincar\'e group the background metric $g(0)_{ij}$ reduces to $\eta_{ij}$ .
Then $R(0)_{ij}=0$, and the covariant derivatives $D(0)_i$
reduce to ordinary partial derivatives. Since we are looking for a solution
given by a perturbation expansion in $\lambda$  we plug in the expressions
\begin{equation}
d_{ij}=d(0)_{ij}+\lambda\, d(1)_{ij}+\lambda^2 d(2)_{ij}+ \cdot\cdot\cdot\;,
\end{equation}
\begin{equation}
a_{ij}=a(0)_{ij}+\lambda\, a(1)_{ij}+\lambda^2 a(2)_{ij}+ \cdot\cdot\cdot\;,
\end{equation}
and then
solve for $d(0)_{ij}$ , $d(1)_{ij}$ , $a(0)_{ij}$ , $a(1)_{ij}$ , etc.

The expansion of (28) in powers of the analytical parameter $\lambda$ reads
\begin{equation}
R_{ij}=R(0)_{ij}+\frac{\lambda}{2}X_{ij}+\frac{\lambda^2}{2}Y_{ij}+\cdot\cdot\cdot\;,
\end{equation}
where
\begin{equation}
X_{ij}\equiv\Box d(0)_{ij}-\partial_k\partial_{(i}d(0)^k_{j)}+\partial_i\partial_jd(0)\;,
\end{equation}
and $d(0)$ is the trace of the matrix $d(0)^i_j$ .

The field equation $R_{ij}=0$ yields one equation for each order in $\lambda$, namely $X_{ij}=0$,  $Y_{ij}=0$, and so forth. The first advantage of our approach is that, $d(0)_{ij}$ being diagonal, the equation $X_{ij}=0$ gets highly simplified. It yields a solution  $d(0)_{ij}$ which is then plugged into $Y_{ij}=0$, from which one works out the solution $d(1)_{ij}$ , and so on.

We now consider the case of an isotropic and static metric, in which case the diagonal metric components further simplify to $d(0)^1_1=d(0)^2_2=d(0)^3_3=\alpha(r)$ and $d(0)^4_4=\beta(r)$ , where $r\equiv\sqrt{\delta_{ij}\,x^ix^j}\,,\;i,j=1,2,3$ .
The first step is solving $X_{ij}=0$ when $i\neq j$ . Then (32) yields
\begin{equation}
\partial_1\partial_2(\alpha(r)+\beta(r))=\partial_1\partial_3(\alpha(r)+\beta(r))=\partial_2\partial_3(\alpha(r)+\beta(r))=0\;,
\end{equation}
which has the trivial solution
\begin{equation}
\alpha(r)+\beta(r)=c+c_1\,r^2\;,
\end{equation}
where $c$ and $c_1$ are constants. For $i=j$ one obtains $\Delta\alpha(r)+2c_1=0$ and $\Delta\beta=0$ when $i=j=1,2,3$ and $i=j=4$ respectively, where $\Delta$ is the 3D Laplacian operator. These equations imply that $\Delta\alpha=\Delta\beta=0$ , and $c_ 1=0$. Then the solutions are $\alpha=-\frac{\omega}{r}+c_2$ and $\beta=\frac{\omega}{r}+c_3$ ; $\omega$, $c_2$ and $c_3$ being new constants. The result $c_1=0$ eliminates the uncomfortable term $\sim r^2$ from the resulting gravitational potential, and the usual boundary conditions at the spatial infinity impose the choice $c_2=c_3=0$, so we finally have
\begin{equation}
\begin{array}{ll}
d(0)^1_1=d(0)^2_2=\!\!\!&d(0)^3_3=-\frac{\omega}{r}\\
      &d(0)^4_4=+\frac{\omega}{r}
\end{array}
\end{equation}
Therefore the contribution of the terms of order $O(\lambda)$ gives rise to a post-Newtonian metric
\begin{equation}
{\rm d}s^2=(1-\lambda\frac{\omega}{r})\,\delta_{ij}\,{\rm d}x^i{\rm d}x^j-(1+\lambda\frac{\omega}{r})\,{\rm d}x^4{\rm d}x^4 \;,
\end{equation}
which allows us to identify $\lambda$ with the gravitational constant. In fact being (33) a $\lambda$-independent equation, $\omega$ becomes an integration constant determined by the central source-mass of the gravitational field.

The study of less symmetrical systems is straightforward. The cylindrical case, as for instance the field  corresponding to a mass distribution along a straight line, gives rise to the well known logarithmical potential. These examples highlight the remarkable simplicity of the first order calculation.

\section{Structure of the general solution at first order in $\lambda$ }

The previous example shows, besides the important simplification,
the remarkable property that the diagonal d.o.f. $d_{ij}$ contribute
to first order in the gravitational constant $\lambda$, whereas the antisymmetric
ones $a_{ij}$ contribute only from $O(\lambda^2)$ on. Thus the basic building block
for the perturbation construction of the solutions is the lowest order
term $d(0)_{ij}$ , the calculation of which relies on solving simple
differential equations. Then we start from (32) and proceed again in two steps.

We first consider the full swing equations $X_{ij}=0$ for $i\neq j$ without any
simplifications. They read
\begin{equation}
\begin{array}{ll}
&\partial_1\partial_2[d(0)^3_3+d(0)^4_4]=0\\
&\partial_1\partial_3[d(0)^2_2+d(0)^4_4]=0\\
&\partial_1\partial_4[d(0)^2_2+d(0)^3_3]=0\\
&\partial_2\partial_3[d(0)^1_1+d(0)^4_4]=0\\
&\partial_2\partial_4[d(0)^1_1+d(0)^3_3]=0\\
&\partial_3\partial_4[d(0)^1_1+d(0)^2_2]=0
\end{array}
\end{equation}
and have the solutions
\begin{equation}
\begin{array}{ll}
&d(0)^3_3+d(0)^4_4=f_1(134)+f_2(234)\\
&d(0)^2_2+d(0)^4_4=m_1(124)+m_2(234)\\
&d(0)^2_2+d(0)^3_3=q_1(124)+q_2(134)\\
&d(0)^1_1+d(0)^4_4=p_1(123)+p_2(234)\\
&d(0)^1_1+d(0)^3_3=s_1(123)+s_2(134)\\
&d(0)^1_1+d(0)^2_2=g_1(123)+g_2(124)
\end{array}
\end{equation}
where the short-hand notation $f_1(134)\equiv f_1(x_1,x_3,x_4)$, etc. has been adopted and $f_1\,,f_2\,,m_1\,,...\,,g_2$ , are arbitrary functions of only three coordinates as displayed in (38).

This large amount of arbitrariness may be reduced by a proper gauge fixing for which we choose the customary harmonic gauge $\partial_i(\sqrt{g}\,g^{ij})=0$, which, at first order in $\lambda$ , yields the differential conditions
$\Box\,d_{ij}=0$ and $\partial_id^i_j=\frac{1}{2}\partial_jd^i_i$ . For a general symmetric matrix the solution to these conditions is undetermined, corresponding to the Lorentz rotations of the four eigenvectors in its spectral expansion. The procedure leading to (23), with $d(0)^i_j$ diagonal, fixes this freedom and has the crucial advantage that the last equation is directly integrable. In fact, it translates into the following conditions on the diagonal metric elements
\begin{equation}
\begin{array}{ll}
&\partial_1[+d(0)^1_1-d(0)^2_2-d(0)^3_3-d(0)^4_4]=0\\
&\partial_2[-d(0)^1_1+d(0)^2_2-d(0)^3_3-d(0)^4_4]=0\\
&\partial_3[-d(0)^1_1-d(0)^2_2+d(0)^3_3-d(0)^4_4]=0\\
&\partial_4[-d(0)^1_1-d(0)^2_2-d(0)^3_3+d(0)^4_4]=0
\end{array}
\end{equation}
with the obvious solutions
\begin{equation}
\begin{array}{ll}
&+d(0)^1_1-d(0)^2_2-d(0)^3_3-d(0)^4_4=h_1(234)\\
&-d(0)^1_1+d(0)^2_2-d(0)^3_3-d(0)^4_4=h_2(134)\\
&-d(0)^1_1-d(0)^2_2+d(0)^3_3-d(0)^4_4=h_3(124)\\
&-d(0)^1_1-d(0)^2_2-d(0)^3_3+d(0)^4_4=h_4(123)
\end{array}
\end{equation}
where the $h_i$ are arbitrary functions with the three coordinates dependence here displayed.
Then the combination of (38) and (40) leaves the much simplified functional structure
\begin{equation}
\begin{array}{ll}
&d(0)^1_1=h_2(134)-h_1(234)+h_4(123)+h_3(124)\\
&d(0)^2_2=-h_2(134)+h_1(234)+h_4(123)+h_3(124)\\
&d(0)^3_3=h_2(134)+h_1(234)+h_4(123)-h_3(124)\\
&d(0)^4_4=h_2(134)+h_1(234)-h_4(123)+h_3(124)
\end{array}
\end{equation}
Now we consider the equations $X_{ij}=0$ for $i=j$. A little work yields
\begin{equation}
\begin{array}{ll}
&\Box\,d(0)^1_1=0\\
&\Box\,d(0)^2_2=0\\
&\Box\,d(0)^3_3=0\\
&\Box\,d(0)^4_4=0
\end{array}
\end{equation}
which, with the constraint (41), finally give
\begin{equation}
\begin{array}{ll}
&(\partial_1\partial_1+\partial_3\partial_3-\partial_4\partial_4)h_2(134)=0\\
&(\partial_2\partial_2+\partial_3\partial_3-\partial_4\partial_4)h_1(234)=0\\
&(\partial_1\partial_1+\partial_2\partial_2-\partial_4\partial_4)h_3(124)=0\\
&\Delta\,h_4(123)=0
\end{array}
\end{equation}

Three remarks are now in order. First, the crucial simplification of the calculations due to
the diagonal form of the symmetric elements $d_{ij}$ . Secondly the physical meaning of the
solutions to (43). In fact, the first three equations are of the hyperbolic type, while the
fourth one is elliptical and yields a static Newtonian potential.

Finally it is interesting to notice that the harmonic gauge gets a new meaning in view of the structure (38).
In fact, the matrix $d_{ij}$ is the result of diagonalizing the symmetric matrix $s_{ij}$
(see (22) and (23)), and then the diagonal elements $d_{ij}$ are the eigenvalues of $s_{ij}$ .
As such, they must be the real roots of a quartic equation, which in turn have the well known parametrization
\begin{equation}
\begin{array}{ll}
&d(0)^1_1=\sigma+\lambda_1+\lambda_2+\lambda_3\\
&d(0)^2_2=\sigma+\lambda_1-\lambda_2-\lambda_3\\
&d(0)^3_3=\sigma-\lambda_1+\lambda_2-\lambda_3\\
&d(0)^4_4=\sigma-\lambda_1-\lambda_2+\lambda_3 \;.
\end{array}
\end{equation}
This parametrization lets finding relationships between the left-hand sides of (38)
such that constraints between the functions $f_1\,,f_2\,,m_1\,,...\,,g_2$ arise, namely
\begin{equation}
\begin{array}{ll}
&[f_1-s_2]+[f_2-m_2]=[g_1-s_1]+[g_2-m_1]\\
&[f_1-q_2]+\,[f_2-p_2]\,=[g_1-p_1]+\,[g_2-q_1]\,.
\end{array}
\end{equation}
The two equations above let eliminating any two of the functions involved in them. What the
harmonic gauge accomplishes, by using (40), is the vanishing of each one of the brackets
in (45), which immediately leads to (41).

\section{Conclusions}

As a consequence of our gauge choice the dynamical variables
get highly simplified, allowing a splitting into a diagonal part and an
antisymmetric part which are not explicit in the metric tensor.
This choice turns out to be most fruitful as long as the diagonal
part contributes to the first order terms, whereas the antisymmetric components
contribute only to higher orders in a characteristic length which acts as a perturbation
expansion parameter and can be identified with the gravitational constant.
This result highlights the deep role that, besides the relevance of
the (local) Lorentz group for the coupling of the fermions, the (local)
translations have as a part of the isometries of a local maximally symmetric
space.

Once an exact solution is known (for instance Schwarzschild's in the static and isotropic case),
in principle it is always possible to derive from it an expansion in the gravitational constant.
On the contrary, our approach provides us with an $ab\;initio$ constructive method for obtaining
such an expansion. Considering that the first order suffices in most applied calculations,
the availability of a simple procedure (for instance, by rendering the harmonic gauge directly
integrable) to obtain the first order terms is of high interest.

In the general case, the field equations at first order are a set of well known differential
equations in D-1 variables, describing a Newtonian potential in the 3D space, and three
hyperbolic equations (popularly referred to as the vibrating membrane equations), defined in
the coordinate 2D subspaces.

The present work obviously sets the basis for further developments. The study of the higher
orders, where the antisymmetric part will give rise to non diagonal metric components, is the
most evident. Among other possibilities, it seems natural to extend the analysis to general
maximally symmetric spaces, which include the cosmological constant.
Work on these topics is in progress.

\section{Acknowledgements}
We acknowledge Prof. A. Fern\'andez-Rañada and J.
Mart\'{\i}n-Mart\'{\i}n, as well as Dr. R. Tresguerres, for useful discussions.
Work partially supported (J. Julve) by the Project MICINN FIS2009-11893.

\end{document}